\documentstyle[12pt]{article}

\begin{document}

\title{
\bf {Time-of-arrival formalism \\for the \\relativistic particle}}
\author{
       { J. Le\'on}\thanks{e-mail: leon@laeff.esa.es} \\
   Laboratorio de Astrof\'{\i}sica Espacial y F\'{\i}sica Fundamental, 
INTA\\
        Ap. 50727, E 28080 MADRID, Spain\\  and \\
         Instituto de Matem\'aticas y F\'{\i}sica Fundamental, CSIC\\
        Serrano 123, E 28001 MADRID, Spain}

\maketitle
\vspace{-10.5cm}
 \hfill LAEFF 96/17
\vspace{9.5cm}
\begin{abstract}
A suitable operator for the time-of-arrival at a detector is defined for 
the free
relativistic particle in 3+1 dimensions. For each detector position, there 
exists a subspace of detected states in
 the Hilbert space of solutions to the Klein
Gordon equation. Orthogonality and 
completeness of the eigenfunctions of the time-of-arrival operator 
apply inside 
this subspace, opening up a standard probabilistic interpretation. 
\end{abstract}
Pacs: 03.65.Bz, 03.65.Ca, 04.60.Ds, 11.30.Cp 

\newpage

\section{Introduction}
 
In non-relativistic dynamics time has a characterization of its own 
which 
distinguishes it sharply from the  space coordinates of configuration 
space. 
However, this difference can be simply removed at the formal level by 
going 
to the parametrized form of dynamics where time is made to depend on a
 parameter
$\tau$ in as much as the coordinates $q^i$ do. One is thus led to deal 
with a
set $(q^i (\tau), t(\tau)) $ in which the identification of time versus 
coordinates appears more as a matter of convention than as a matter of 
significance from the point of view of the dynamical system under study.
Even though, time still keeps a particular role from the physical point 
of
view. Time is experienced by the observer as well as by the system. This 
is more 
evident in the transition to quantum mechanics, where time
-as opposed to position- can not be viewed as a property of the system 
under scrutiny.

There is a way out from this situation as shown in ref~\cite{grot}, 
whose authors 
show how to deal with and solve the question {\sl at what time?} in 
quantum mechanics
 in one space dimension by introducing a suitable time operator, and 
obtaining the 
associated time representation. The outcome is the emergence of a 
$x\leftrightarrow t$
 equivalence in quantum mechanics in much the same way as there is one in
 classical mechanics.
The question {\sl at what time?} joins the question {\sl at what 
position?} as
answerable not only experimentally, but also within the realm of the 
quantum mechanical
formalism.  

In special relativity time is obviously  $q^0$, and it seems the 
question {\sl at what
time?} would be addressed in relativistic quantum mechanics in a simple
 and direct way: 
explicit covariance should rule
the presence of $q^0$ along with the space components $q^i$ to form a 
Minkowski space
fourvector $q^\mu$. There should be no telling difference between the 
time and the
space components of $q$, mainly taking into account that -in contrast to 
 the non 
relativistic case- they get entangled by Lorentz transformations. One 
could be led to 
believe in the existence of a space-time position operator, a four-vector,
 whose 
components should transform covariantly under the Lorentz group. This 
object should address
simultaneously the two questions {\sl when?} and {\sl where?} seemingly
 unrelated in the 
non relativistic case. It is well known that this object has never been
 constructed.
In the instant form of dynamics, i.e. refering the operators to their 
values at some 
instant of time, one can employ a three-vector operator -the position 
operator~\cite{newt}- to 
answer the question {\sl where?} This operator not only lacks explicit 
covariance,
it also lacks a time component. The cause of these defficiencies can be 
traced back~\cite{har}
 to the
reparametrization invariance of the action of the relativistic particle
\begin{equation}
{\cal S}= m \int d\tau \sqrt{\dot{q}^2} \label{1}
\end{equation}
which translates into evolution (along $\tau$) generated by a Hamiltonian 
$H=p^2 -m^2=0$. Since the Hamiltonian is constrained to vanish, the $\tau$
evolution is a gauge transformation. In 
the canonical 
approach  one  chooses a solution
to the constraint, i.e. by putting $p^0=\sqrt{\vec{p}^2 +m^2}$, and 
``fixes the gauge''
by setting the evolution parameter  to be the physical time. 
A priori there is no room left for the
question {\sl when?} as there is no freedom left for a time operator 
differing from 
the time parameter $q^0$. This is a bonus from another point of view: 
demoting $q^0$
to the role of a parameter one evades the difficulty of a Hamiltonian
 unbounded from 
below in the same way as in the non-relativistic case. The lack of 
positivity of
the density $j^0$ of the solutions of the Klein Gordon equation also 
plays a role here.
It brings about particle-antiparticle pairs, etc. and the untenability 
of the one
particle interpretation. From here on, the true variables are field 
configurations,
to whom $q^0$, along with the space coordinates $q^i$, are mere 
parameters. However, 
the case of the relativistic particle we are analyzing here is of 
intrinsic interest; it 
serves to
set up the basis for the particle interpretation of quantum field theory,
 and also
 as a guideline to use~\cite{teit} in the construction of the quantum 
formalism of the 
gravitational field. Analyzing issues of time for the relativistic
 particle may
prove valuable in transforming that formalism in a theory or, at least,
 may throw
some light on the issues of time in quantum gravity~\cite{is}. This 
paper focuses on the
relativistic particle. In Sect. 2 we summarize the results of the 
canonical formalism, 
In section 3 we generalize the treatment of Ref.~\cite{grot} to the 
free relativistic
particle, Sect. 4 contains the generalization to three space dimensions 
and Sect. 5 is
devoted to questions of orthogonality and completeness. Finally, in  
section 6 we discuss some
issues raised by the interpretation of the formalism and some 
speculations about the
applicability to quantum gravity. 
 
\section{Canonical formalism}
Here we will focus our attention onto the physical Hilbert space 
${\cal H}_{KG}$
 of the positive energy solutions  $\psi (x)$ for the Klein Gordon 
equation~\cite{hal}, 
with 
the understanding that negative energies will be reinterpreted in terms 
of antiparticles. In configuration space where the Klein Gordon equation
reads $(\Box + m^2)\psi (x)=0$,
the positive energy solutions are of the form:
\begin{eqnarray}
\psi (x)&=&(2\pi)^{-3/2} \int d^4 k e^{-ikx} \delta(k^2 -m^2) \theta 
(k^0)
\Psi (k) \nonumber \\
&=&(2\pi)^{-3/2} \int \frac{d^3 k}{2 \omega(k)} e^{-i(\omega (k) x^0 
-\vec{k}\vec{x})}
\Psi(\vec{k})  \label{2}
\end{eqnarray}
with a scalar product:
\begin{equation}
(\phi,\psi)=i\int d^3 x (\phi^*\partial_t \psi-\partial_t \phi^* \psi)
           =\int \frac{d^3 k}{2 \omega(k)} \Phi^*(\vec{k}) \Psi(\vec{k}),
\label{3}
\end{equation}
where $\omega (k)=\sqrt{\vec{k}^2 +m^2}$. We will follow the conventions 
of~\cite{schwe} denoting with uppercase 
letters the wave 
functions in momentum space, leaving the lower case for configuration 
space 
functions.

To answer the question ``What is the probability of finding the particle
 at 
the point $\vec{x}$ at time $x^0$?'' with the above scalar product, we 
need to
find a Hermitian position operator and find its eigenfunctions 
$\psi_{\vec{x},x^0}$.
Then, the probability amplitude in $\vec{x}$ for finding  a particle
 at $\vec{x}$ at time $x^0=q^0$ is $(\psi_{\vec{x},x^0},\phi)$, 
where $\phi(q)$ is the wave function giving the state of the particle. 
As shown by
 Newton and 
Wigner~\cite{newt} the position operator is
\begin{equation}
\vec{Q}=i\vec{\nabla}_p -\frac{i \vec{p}}{2 \omega^2(p)} \label{7}
\end{equation}
In our notation,  $k$ will represent $p$ in momentum space,  while 
 $Q$'s and $p$'s will denote operators, unless specified otherwise by 
the word ``classically'', in which case they will denote classical 
dynamical variables . The eigenstate of the 
position operator localized at the point $\vec{x}$
 at $t=0$ is
\begin{equation}
\Psi_{\vec{x},0}(\vec{k})=(2\pi)^{-3/2} \sqrt{2\omega(k)} e^{-i\vec{k}
\vec{x}}\label{8}
\end{equation}
In general, given a particle in the state $\Phi(\vec{k})$ at $t=0$, the 
probability 
amplitude to find it at the position $\vec{x}$ at $t=0$ is given by
\begin{equation}
(\Psi_{\vec{x},0},\phi)=(2\pi)^{-3/2}\int\frac{d^3 k}{2 \omega(k)}
e^{i\vec{k}\vec{x}} \sqrt{2 \omega(k)}\Phi(\vec{k}) \label{9}
\end{equation}

The components of the position operator are in involution and commute 
canonically 
with the momenta
\begin{equation}
[Q^i,Q^j]=0,\, [Q^i,p^j]=i \delta^{ij} \label{10}
\end{equation}
under rotations and space translations $\vec{Q}$ behaves as a three 
vector. It also 
evolves like the position of a particle should do, namely
\begin{equation}
\frac{d\vec{Q}}{dt}=i[\sqrt{\vec{p}^2 +m^2},\vec{Q}]=\frac{\vec{p}}
{\omega(p)} 
\label{11}
\end{equation}
The Heisenberg position operator at time $t$ can be obtained by 
integrating this equation
\begin{equation}
\vec{Q}(t)=\vec{Q}+\frac{\vec{p}}{\omega(p)} t \label{12}
\end{equation}
 We now 
would like to 
invert this equation to get an operator for the time-of-arrival of the 
relativistic 
particle following the proposal of~\cite{grot}. 

\section{Time-of-arrival in one space dimension}
The special role played by time has been the source of controversy since
the early days of quantum mechanics. The search of the various time
operators and the analysis of the associated time-energy uncertainity 
relations was the subject of a number of works (see the bibliography in 
ref.~\cite{muga}), whose outcome was that 
quantum mechanics can not accomodate a time-of-arrival operator. This has been
 refuted recently in ref.~\cite{muga} where, in addition, an average value
for this quantity is explicitly obtained for one space dimension in terms
of the current density of the particle. This  is framed in a wealth of
recent works devoted to the issue of time in quantum mechanics 
-see ref.~\cite{muga2} and the bibliography contained therein- 
with
special emphasis on the tunnelling times, a question of fundamental and
practical implications. Here, we are interested in the characterization of the
time-of-arrival as one of the properties of the system under study as in 
ref.~\cite{grot}, in other words, we need to go
one step further and to obtain
an associated operator to be able to analyze and give
an interpretation to  this property in the quantum formalism. This is 
necessary for our results to be of value for the quantum formalism of the
gravitational field where, as said in the introduction, time has to be
considered as a property of the system under study.
For the sake of simplicity and also
to connect with the non-relativistic one-dimensional case studied 
in~\cite{grot} we
begin by considering the case of one space dimension. Then we can 
rewrite (\ref{12}) as
\begin{equation}
Q(t)=Q+\frac{p}{\omega(p)} t \label{14}
\end{equation}
and the time-of-arrival at the position $Q(t)=X$ would be given by a suitable
 ordering of
the operator
\begin{equation}
Q^0(X)\simeq (X-Q)\frac{\omega(p)}{p} \label{15}
\end{equation}
where the simbol $\simeq$ is employed to mean equal apart from ordering.
Now, $Q^0(X)$
can be given simply in a form that goes to the operator ${T}(X)$ 
of~\cite{grot} in the
non-relativistic limit:
\begin{equation}
Q^0(X)=e^{-ipX}\sqrt{\frac{\omega(p)}{p}}(-i\frac{d}{dp}+\frac{ip}
{2 \omega^2(p)})
\sqrt{\frac{\omega(p)}{p}} e^{ipX} \label{16}
\end{equation}
The eigenfunctions of this operator
\begin{equation}
Q^0(X) \Psi_{T,X}(k)=T \Psi_{T,X}(k) \label{17}
\end{equation}
are given by
\begin{equation}
\Psi_{T,X}(k)=\alpha \sqrt{k} e^{i(\omega(k)T-kX)}\label{18}
\end{equation}
where $\alpha$ is a normalization factor.
Multiplying by the phase factor $\exp{(-imT)}$, these functions 
give the eigenfunctions of~\cite{grot} in the non-relativistic 
limit. We will not make distintions
 between right ($k>0$) and left moving ($k<0$)  particles here, as these
 have a meaning for 
one space dimension  only and we want to study the 3-D case, where 
opposite directions can be connected  continuously.

\section{Three space dimensions}
A new feature appears in three space dimensions that was not present in 
the case 
studied above. The  space of ``detected'' states is a subspace of  
the Hilbert
space ${\cal H}_{KG}$ of positive energy solutions to the Klein Gordon 
equation. 
This comes about because in the 3-D case the evolution equations that we
 have to invert
 to obtain the time-of-arrival is the set (\ref{12}) of three equations 
depending on
a unique parameter $t$. To be compatible, they have to satisfy the 
constraint
\begin{equation}
\vec{\cal C}= (\vec{Q}-\vec{X}) \wedge \vec{p}=0 \label{19}
\end{equation}
where the ``point-of-arrival'' $\vec{X}$ plays the role of a parameter. 
Classically, these 
constraints mean that the angular momentum of the particle is 
$\vec{X}\wedge\vec{p}$,
so that $\vec{X}$ is a point in the particle trajectory, or simply  that 
the angular momentum about $\vec{X}$ is zero. In quantum 
mechanics there are 
obstructions to imposing simultaneous values to different components of 
the angular
momentum. At first sight, the best one can do is to constrain 
$\vec{L}^2$ and a component 
of the angular
momentum, say $L_3$, to have definite values given from 
$\vec{X}\wedge \vec{p}$. However,
this is not the case here, as we are equating the components of the angular
 momentum to an 
operator $\vec{X}\wedge\vec{p}$, in such a way that the constraints form
 a first class
system.  Classically, Eq. (\ref{19}) plays the role of a set of first class 
constraints in the hamiltonian formalism that we have to quantize following
 the method of Dirac. Now, the total hamiltonian is
\begin{equation}
H=\sqrt{\vec{p}^2+m^2} +\lambda_a {\cal C}_a \label{20}
\end{equation}
where
\begin{equation}
{\cal C}_a=\epsilon_{abc} (Q-X)_b p_c \label{21}
\end{equation}
and the $p$'s and $Q$'s are the dynamical variables to become operators
after quantization. 
It is straightforward to show that
\begin{equation}
\{{\cal C}_a,{\cal C}_b\}=\epsilon_{abc}{\cal C}_c, \, \{{\cal C}_a,H\}=
\epsilon_{abc}
\lambda_b {\cal C}_c \label{22}
\end{equation}
Therefore, we have a true first class system, a different one for each 
vector $\vec{X}$.

There seems to  be additional difficulties in that the eigenvalues of 
$\vec{L}^2$ and $L_3$ are integer
numbers while the constraint will assign to them a continuous spectrum. 
Actually, this is not the case~\cite{marolf} because, even if the 
constraint can be
written in the form $\vec{L} = \vec{X}\wedge \vec{p}$, this will not
hold as an operator equation, nor the states on wich it will be 
satisfied will be  eigenstates of neither $L_i$ nor $\vec{X}\wedge
\vec{p}$. 
 Now, the detected  subspace
${\cal H}_{KG}^{(X)}$ can be given simply as that spanned by the 
functions 
$\Psi^{(X)}(\vec{k})$ of the form

\begin{equation}
{\cal H}_{KG}^{(X)}=\left\{\Psi^{(X)}(\vec{k})=e^{-i\vec{k}\vec{X}}
\Psi(k,\vec{X})
\right\}\label{a}
\end{equation}
where $\Psi(k, \vec{X})$ represents an arbitrary function of the modulus
 of $\vec{k}$ and of
$\vec{X}$.  If we now require  invariance under 
translations, we have to drop 
the dependence of $\Psi(k,\vec{X})$ on $\vec{X}$. In this case we can 
say that the Hilbert
space ${\cal H}_{KG}^{(X)}$ is obtained from  ${\cal H}_{KG}^{(0)}$ by
 a translation of
amount $\vec{X}$.

We are now prepared to study  $Q^0(\vec{X})$, the time-of-arrival at
a point $\vec{X}$ in the 3-D space. Classically, it is given by inverting
the equation of motion:
\begin{equation}
Q^0(\vec{X})=\frac{\omega(p)}{\vec{p}^2}\, (\vec{Q}-\vec{X})\cdot\vec{p},
\end{equation}
which is a first class dynamical variable 
$\{ Q^0(\vec{X}), {\cal C}_a \} =0$.
In the Hilbert space 
${\cal H}^{(X)}_{KG}$ 
 the operator equation of motion has to be rewritten with $t$ replaced
by the operator $Q^0(\vec{X})$ and $\vec{Q}(t)$ by the detector's position
$\vec{X}$
\begin{equation}
\vec{X}-\vec{Q}-\frac{\vec{p}}{\omega(p)} Q^0(\vec{X})
= 0 \label{27}
\end{equation}
It should be an identity, with the operator $Q^0$ being such to 
annihilate the left hand 
side.  By vector product of the above equation by $\vec{p}$ we obtain 
the 
constraints that are already satisfied in the detected subspace. Scalar 
product by $\vec{p}$
gives
\begin{equation}
\vec{p}\vec{X}-\vec{p}\vec{Q}-\frac{p^2}{\omega(p)}Q^0(\vec{X})
= 0 \label{b}
\end{equation}
Putting
\begin{equation}
Q^0(\vec{X})=e^{-i\vec{p}\vec{X}} Q^0e^{i\vec{p}\vec{X}} \label{28}
\end{equation}
the previous equation reduces to
\begin{equation}
-i\frac{d}{dp}+\frac{ip}{2\omega^2(p)}-\frac{p}{\omega(p)}Q^0
= 0 \label{29}
\end{equation}
Observe how, when acting on the detected subspace, Eq. (\ref{27}) 
reduces effectively to
only the one-dimensional equation (\ref{29}).
One would be tempted to solve it with the ordering chosen in (\ref{16}),
 with 
eigenfunctions similar to (\ref{18}). This choice would not do, as the 
norm of these states
 would be badly divergent in three dimensional space. What we need are 
eigenstates with 
higher negative powers of $k$ than in (\ref{18}). This can be achieved 
by choosing a different
 ordering for the operator. Tentatively we put
\begin{equation}
Q^0=\sqrt{\omega(p)}\frac{1}{p^{n+1}}\left(-i\frac{d}{dp}+\frac{ip}
{2\omega^2(p)}\right)
p^n \sqrt{\omega(p)} \label{30}
\end{equation}
with this choice we get for the eigenfunction of (\ref{28}) with 
eigenvalue $T$ the expression
\begin{equation}
\Psi_T^{(X)}(\vec{k})=\frac{1}{2 \pi k^n} e^{i(\omega(k)T-\vec{k}
\vec{X})} \label{31}
\end{equation}
where we have chosen some arbitrary fixed $\vec{X}$. We now choose 
$n$ such that the scalar product be well behaved
\begin{eqnarray}
(\psi^{(X)}_T,\psi^{X}_{T'})&=&\int\frac{d^3 k}{2\omega(k)} 
\Psi^{(X)^*}_T(\vec{k})
\Psi^{(X)}_{T'}(\vec{k})\nonumber\\
&=&(2\pi)^{-1} \int_0^{\infty} dk\frac{k^{2(1-n)}}{\omega(k)}
e^{i\omega(k)(T'-T)} \label{32}
\end{eqnarray}
We see that the eigenfunctions are not orthogonal. We will address 
this problem in the next section. Now, we focus on the last integral, 
which strongly suggest the 
choice $n=1/2$. In the general case of $d$ space dimensions we would chose
$n=(d-2)/2$, to make the measure of the integral equal to $d\omega$. 
Finally, in our case we have:
\begin{eqnarray}
Q^0& =&\sqrt{\omega(p)}p^{-3/2}\left(-i\frac{d}{dp}+\frac{ip}
{2\omega^2(p)}\right)
          p^{1/2}\sqrt{\omega(p)} \nonumber \\
\Psi^{(X)}_T(\vec{k})&=&(2\pi )^{-1}k^{-1/2} \ 
e^{i(\omega(k)T-\vec{k}\vec{X}}) 
\label{33} \\
(\psi^{(X)}_T,\psi^{(X)}_{T'})&=&(2\pi)^{-1}\int_m^{\infty} d\omega 
e^{i\omega(k)(T'-T)} \nonumber
\end{eqnarray}
 If there
is any doubt left in that the right choice is $n=1/2$, one can check 
that this value gives the 
unique ordering that makes the operator $Q^0$ Hermitian, 
$(\phi,Q^0 \psi)=(Q^0 \phi,\psi)$.

\section{Orthonormalization and completeness}
The eigenfunctions of (\ref{33}) are not yet orthogonal. However the 
above 
scalar product
is an appropriate expression for the Marolf's orthogonalization 
re\-ci\-pe~\cite{grot}. It 
is based in the physical observation that for vanishing momentum the 
particle either never
reaches the detector, or sits in it forever. To deal with this 
situation, Marolf proposed a regularization prescription for the 
time-of-arrival
 operator that ``avoids'' zero momentum particles. The 
procedure to follow is less obvious here than in the 1-D 
non-relativistic case, due to
the more complex structure of the operator. We first present the 
appropriate prescription
for arbitrary $n$, coming back to $n=1/2$ at the end of the calculation,
 to show that only with
this value the procedure gives orthogonal eigenfunctions in three space 
dimensions. First,
we rewrite $Q^0$ in the momentum representation  as 
\begin{equation}
Q^0 = -i \omega(k) \frac{1}{k^{n+1/2}\sqrt{k}} \frac{d}{dk}
\frac{k^{n+1/2}}{\sqrt{k}}
\label{34}
\end{equation}
which we regularize as follows
\begin{equation}
Q^0 = -i \omega(k) \sqrt{f(k)}\frac{1}{k^{n+1/2}} \frac{d}{dk} 
k^{n+1/2}\sqrt{f(k)},
\label{35}
\end{equation}
and where $f$ is the same as in~\cite{grot}
\begin{equation}
f(k)= \left\{ \begin{array}{ll} \frac{1}{k} &\mbox{for}\ k>\epsilon \\
                               \epsilon^{-2} k &\mbox{for}\ k<\epsilon
\end{array}
\right. \label{36}
\end{equation}
The eigenfunctions $\Psi_T^{(X)}(\vec{k})$ corresponding to this 
operator are of the form:
\begin{equation}
\Psi^{(X)}_T(\vec{k}) =\frac{1}{2\pi}\frac{e^{i(Z(k)T-\vec{k}\vec{X})}}
{k^{n+1/2}\sqrt{f(k)}},\, \,
Z(k)=\int_\epsilon^k\frac{dk'}{\omega(k')f(k')}, \label{37}
\end{equation}
and the orthogonality condition reads
\begin{equation}
(\psi^{(X)}_T,\psi^{(X)}_{T'})=(2\pi)^{-2}\int\frac{d^3k}
{2\omega(k)f(k)}\frac{1}{k^{2n+1}}
e^{iZ(k)(T'-T),}  \label{38}
\end{equation}
For the case $n=1/2$ one gets
\begin{equation}
(\psi_T^{(X)},\psi_{T'}^{(X)})=(2\pi)^{-1}\int_{Z_{min}}^{Z_{max}} dZ 
e^{iZ(T'-T)}=
\delta(T-T'), \label{39}
\end{equation}
as the coordinate $Z$ goes from $-\infty$ to $0$ as $k$ goes from $0$ to
 $\epsilon$,
and from $0$ to $\infty$ as $k$ goes from $\epsilon$ to $\infty$.  $Z$
 and $T$ form a 
pair of ``conjugate'' variables in the subspaces ${\cal H}_{KG}^{(X)}$. 
This can be seen
from (\ref{39}) and the associated completeness relation 
\begin{equation}
\int_{-\infty}^{+\infty}dT \Psi^{(X)}_T(\vec{k})\Psi^{(X)^*}_T(\vec{k}')=
\frac{1}{2\pi k^2 f(k)}\delta(Z(k)-Z(k')) \ e^{-i(\vec{k}-\vec{k}')
\vec{X}}\label{40}
\end{equation}
The weird expression on the rhs is exactly what is needed to form a 
completeness relation
in the detected subspace. For any function 
$\Phi^{(X)}\in{\cal H}_{KG}^{(X)}$
\begin{equation}
\int\frac{d^3 k'}{2\omega(k')}\left\{\int_{-\infty}^{+\infty}dT 
\Psi^{(X)}_T(\vec{k})
\Psi^{(X)^*}_T(\vec{k}')\right\}\Phi^{(X)}(\vec{k}')=\Phi^{(X)}
(\vec{k})\label{41}
\end{equation}
as should be expected. In addition, using the expressions (\ref{35}) for
 $Q^0$ and 
(\ref{37}) for $Z$, the following commutation rule is derived
\begin{equation}
[Q^0,Z]=-i \label{42}
\end{equation}
The spectral support of both $Q^0$ and $Z$ is the whole real line, so
 that no difficulties
arise from the Stone - Von Neumann theorem with (\ref{42}) as would be 
the case were it to involve 
$\omega$ instead of $Z$.
Finally, a comment on the relation between the time and the position 
operators is 
in order: The eigenstates of $\vec{Q}$ with eigenvalue $\vec{X}$ 
(\ref{8}) belong 
to the detected subspace ${\cal H}_{KG}^{(X)}$. However, it is not 
possible to determine
simultaneously both
the position (or the momentum) and the time-of-arrival due to the fact
 that the 
corresponding operators do not commute.

\section{Interpretation}
The results obtained so far indicate that the 
operator formalism associated to the time-of-arrival at a point works 
to fit the quantum mechanical rules. Accordingly, one can 
interpret  it in a novel but standard way as was done on physical
grounds in ref~\cite{grot} for one space dimension. Here, we will show 
that the formalism provides the tools with which to build the
quantum mechanical interpretation to be given to the time-of-arrival
operator. In other words, that it provides the mathematical framework
sufficient to define the time-of-arrival properties of the particle
and associate to them definite probabilities. For definiteness, we assume
that we are analyzing the time-of-arrival at the point $X$. First, we split
the Hilbert space $\cal H$ of states into never detected ${\cal H}_{ND}$
and detected subspaces ${\cal H}_D$; obviously ${\cal H}={\cal H}_{D}
\oplus {\cal H}_{ND}$. Also, from the discussion in Section 4, we know
that  ${\cal H}_{D}={\cal H}^{(X)}$. This will be the Hilbert space
appropriate to the analysis. In ${\cal H}^{(X)}$ we have defined the 
(regularized) Hermitian operator $Q^0 (\vec{X})$, whose spectrum is
$T\in {\cal R}$, the set of observable times-of-arrival at the point $X$.
Having solved the eigenvalue problem for  $Q^0 (\vec{X})$, we obtained a
complete and orthogonal set of eigenfunctions $\psi^{(X)}_T (\vec{k})=
<\vec{k}|T, \vec{X}>$ in the momentum representation. From them, we
can define the set of elementary projectors 
$\{\Pi^{(X)}_T,T\in {\cal R}\}$ where
\begin{equation}
\Pi^{(X)}_T=|T,\vec{X}><T,\vec{X}| \label{60}
\end{equation}
They generate a boolean algebra ${\cal B}$ with the properties
\begin{equation}
 \Pi^{(X)\dagger}_T=\Pi^{(X)}_T, \, \,
\Pi^{(X)}_T \Pi^{(X)}_{T'}=\delta (T-T')\Pi^{(X)}_T \label{61}
\end{equation}
 To each elementary projector there corresponds an event
($\Pi^{(X)}_T \leftrightarrow$ arrival at time $T$).
Given any two projectors $\Pi, \Pi' \in {\cal B}$ the
meet ({\sl and}) and join ({\sl or}) operations are defined as usual by
\begin{equation}
\Pi \wedge \Pi' =\Pi \Pi', \, \, 
\Pi \vee \Pi'=\Pi +\Pi' -\Pi \Pi' \label{65}
\end{equation}
where the notation corresponding to a finite dimensional Boole algebra
has been displayed for simplicity. 
Statements  will in general be of the form
($Q^0 (\vec{X}),T_1 <T<T_2$), i.e. the particle arrives at $X$ in the 
interval ($T_1,T_2$). Associated to them there will be projectors built 
by the joining of elementary projectors of the algebra 
\begin{equation}
\Pi^{(X)}(T_1,T_2)=\int_{T_1}^{T_2} dT\ \Pi^{(X)}_T  \label{62}
\end{equation}
with matrix elements
\begin{equation}
<T,\vec{X}|\Pi^{(X)}(T_1,T_2)|T',\vec{X}>=\delta(T-T')
\theta (T_2-T)\theta (T-T_1)\label{63}
\end{equation}
Finally, the algebra has to provide a decomposition of the
identity suitable for the analysis of the properties of the observable
under discussion, i.e.
\begin{equation}
\Pi^{(X)}\equiv \int_{-\infty}^{+\infty} dT \Pi_T^{(X)}=1 \label{64}
\end{equation}
which is valid in ${\cal H}^{(X)}$ due to (\ref{41}), with the obvious 
meaning that an arbitrary state of  ${\cal H}^{(X)}$ will not escape
from detection. When acting on states belonging to Hilbert spaces
larger than  ${\cal H}^{(X)}$ the value of $\Pi^{(X)}$ will be smaller 
than one.  

The complement of the statement $ \Pi^{(X)}(T_1,T_2)$, i.e. the particle 
 arrives at $X$ at a time outside the interval $(T_1,T_2)$ will be 
given by the projector $\Pi^{(X)}-\Pi^{(X)}(T_1,T_2)$. 
 In the case that the state of the particle belongs to  ${\cal H}^{(X)}$ 
the  complement  gives simply $1-\Pi^{(X)}(T_1,T_2)$. The statement
 that there are states that escape from detection, absolutely when 
their projection on the detected subspace vanishes, or partially when
they do not belong to  ${\cal H}^{(X)}$ but have a finite projection on it,
is given by the projector $1-\Pi^{(X)}$. Finally, joining this last to the 
complement, gives the negative statement $1-\Pi^{(X)}(T_1,T_2)$,
i.e. the particle does not arrive at $X$ in the interval $(T_1,T_2)$. The 
fact that the negation and the complement may  differ is a consequence
of the incomplete character of the spectral decomposition of the 
time-of-arrival operator ($\Pi^{(X)}<1$). This could be avoided by working
inside ${\cal H}^{(X)}$ only, but this is too small to be of practical 
interest, consisting only of spherical waves about $X$.

We can now assign probabilities to the statements represented by the 
 projectors of the algebra ${\cal B}$. Given an arbitrary normalized state
$\Phi$ of the physical Hilbert space, the probability (in time) of arriving
during the interval $(T_1, T_2)$ at the position $\vec{X}$, $P_T^{(X)}(\Phi)$
is given by
\begin{equation}
P_{(T_1,T_2)}^{(X)}(\Phi)=\int_{T_1}^{T_2} dT\, |<T,\vec{X}|\Phi>|^2
\label{70}
\end{equation}
An arbitrary state $\Phi$ does not need to be in ${\cal H}_D$, but in 
general will have a finite projection on it. Accordingly, we can define 
the probability of being ever detected at $\vec{X}$ by
\begin{equation} 
P^{(X)}(\Phi)=\int_{-\infty}^{+\infty} dT\, |<T,\vec{X}|\Phi>|^2
\label{71}
\end{equation}
This will be equal to one for normalized states in ${\cal H}_D$, as can be
 obtained from (\ref{41}). For states not in ${\cal H}_D$ this describes the 
case of states that classically would never be 
detected at the position $\vec{X}$, but quantum mechanically have a
-less than one, but finite- probability for (ever) being detected at that 
point. Consider for example the ideal situation in which we place a detector
along the {\sl ox} axis at $\vec{X}=(x,0,0)$, and prepare at $t=0$ a 
gaussian wave packet centered at the origin, with mean momentum slightly
off the {\sl ox} axis $<\vec{k}>=(k_0 \sin\theta,0,k_0 \cos\theta)$.
We consider the uncertainities in position and momentum to be such that
wave packet and detector are well separated at $t=0$, and the cone of
flight of the particle $(\delta\theta\sim \Delta k/k)$ misses the detector.
Even in this case, there
will be a small probability for the particle being ever detected at 
$\vec{X}$; it is given by $P^{(X)}(\Phi)$. The probability of being
detected during the interval $(T_1,T_2)$ will be given by
$P_{(T_1,T_2)}^{(X)}(\Phi)$, while the average value of the time-of-arrival
operator will be
\begin{equation}
<Q^0(\vec{X})>=\frac{\int_{-\infty}^{+\infty} dT\, T |<T,\vec{X}|\Phi>|^2}
{\int_{-\infty}^{+\infty} dT\, |<T,\vec{X}|\Phi>|^2}
\label{73}
\end{equation}
This is a conditional average value, i.e. it makes sense only in the case
when the particle is ever detected. Speaking about the value of the 
time-of-arrival in the other case is a logical contradiction,
undefined mathematically, as in this case $<T,\vec{X}|\Phi>=0$.

 The question of the time-of-arrival still deserves further 
clarification in quantum mechanics. We have outlined the 
mathematical framework whose existence allows for the asignment of
probabilities to its different statements and for the use of logic
to make inferences. In doing this, we are implicitly considering the
existence of measurement devices (detectors in this case) which will
function almost ideally, without introducing serious disturbances in 
the experimental results, so that the logical outcomes can be compared
straightforwardly with the actual results. The existence of such detectors
goes beyond the scope of the present work, which only deals with the 
formalism and its interpretation. This is a question common to
this (distributions in time), and the usual (distributions in space)
formulations of Quantum Mechanics, and we can think that what is 
applicable there is also applicable here. Other serious issue, of actual
interest for its practical implications, is the inclusion of interactions
in the formalism. For instance, How will the gravity field of the Earth
modify the distribution of times-of-arrival as measured on the laboratory?
This is of interest as there are experiments based on the production of
a time-of-flight spectrum against the force of gravity. Another question 
is that of the time-of-arrival at a detector of a particle after traversing
 a barrier by quantum tunnelling. There is no classical analog to this 
situation. Therefore the method presented here will be useless to address
this problem, which calls for a completely quantum mechanical approach.  
There is a long list of pending questions worth of further research. Here, we
turn to one of the motivations of this work:   
 using the
relativistic particle as a guideline to learn about  time 
in quantum gravity. In principle, it would be plausible to think of 
the space part of 
the metric as playing a role similar to that of the detector position. 
Then, constraints
restricting the detected Hilbert space as in (\ref{19}) are likely to 
appear. 
Were this the case, the comparison would be among different possible 
initial states 
(of the Universe (?)), and the
subject of comparison the time employed by these states to -or the 
probability of- 
``evolve''~\cite{evo} to a definite space metric. All this is highly 
speculative and
 object of
further research. First of all, it is not even clear the mere existence 
of a suitable
classical scheme from which to derive a time operator in the general 
case.  
\section*{Acknowledgments}
The author would like to thank to R. S. Tate for his comments which
have improved so much the final version of this paper, and to D.
Marolf for helpful correspondence. He also 
 thanks to F. Barbero, F. Gaioli, E. Garc\'{\i}a Alvarez and
D. Hochberg for useful discussions, and to R. Tresguerres,
J. Julve, A. Tiemblo and F. J. de Urries for their interest in this
work.

\end{document}